\documentclass[pre,aps,twocolumn,showpacs,preprintnumbers,superscriptaddress]{revtex4}

\usepackage{graphicx}
\usepackage{dcolumn}
\usepackage{bm}
\usepackage{amssymb}
\usepackage{pifont}
\usepackage{amsmath}
\usepackage{times}
\usepackage[nooneline]{subfigure}

\begin{document}

\title{Energy dissipation statistics in the random fuse model}

\author{Clara B.\ Picallo} \email{picallo@ifca.unican.es}

\affiliation{Instituto de F\'{\i}sica de Cantabria (IFCA), CSIC--UC, E-39005
Santander, Spain}

\affiliation{Departamento de F{\'\i}sica Moderna, Universidad de Cantabria,
Avda. Los Castros, E-39005 Santander, Spain}

\author{Juan M.\ L{\'o}pez} \email{lopez@ifca.unican.es}

\affiliation{Instituto de F\'{\i}sica de Cantabria (IFCA), CSIC--UC, E-39005
Santander, Spain}

\date{\today}

\begin{abstract}
We study the statistics of the dissipated energy in the two-dimensional random fuse model
for fracture under different imposed strain conditions. By means of extensive numerical
simulations we compare different ways to compute the dissipated energy. In the case of a
infinitely slow driving rate (quasi-static model) we find that the probability
distribution of the released energy shows two different scaling regions separated by a
sharp energy crossover. At low energies, the probability of having an event of energy $E$
decays as $\sim E^{-1/2}$, which is robust and independent of the energy quantifier used
(or lattice type). At high energies fluctuations dominate the energy distribution leading
to a crossover to a different scaling regime, $\sim E^{-2.75}$, whenever the released
energy is computed over the whole system. On the contrary, strong finite-size effects are
observed if we only consider the energy dissipated at microfractures. In a different
numerical experiment the quasi-static dynamics condition is relaxed, so that the system is
driven at finite strain load rates, and we find that the energy distribution decays as
$\mathcal{P} (E) \sim E^{-1}$ for all the energy range.
\end{abstract}

\pacs{46.50.+a, 62.20.Mk}

\maketitle

\section{Introduction} Acoustic emission (AE) in fracture is an example of a broader
phenomenon known as `crackling noise'~\cite{sethna}. A system crackles in response to an
external stimulus leading to energy dissipation in the form of avalanches of events with
no characteristic size. Examples can be found in volcanic rocks~\cite{diodati}, crumpling
paper~\cite{houle}, superconductors~\cite{field}, disordered magnets~\cite{durin}, or
plastic deformation of materials~\cite{carmen}. The release of acoustic energy in fracture
experiments is related to irreversible processes, like creation of microcracks and
deformation.

AE experiments in stressed materials typically give a power-law distribution of the
dissipated energy, $\mathcal{P}(E) \sim E^{-\beta}$, as the system is slowly driven
towards catastrophic failure. This can be seen as a small-scale analog of the
Gutenberg-Richter law for earthquakes. The exponent $\beta$ is found to depend on the
material and in the literature one can find $\beta \approx 1.3$ in synthetic
plaster~\cite{petri}, $\beta \approx 1.25$ in paper~\cite{alava02}, $\beta \approx 2.0$ in
fiberglass and $\beta \approx 1.51$ in wood~\cite{garci1,garci2,garci3}, $\beta \approx
1.5$ in cellular glass~\cite{maes}, and $\beta \approx 1.8$ in paper peeling
experiments~\cite{salminen06}. This spread of values indicates that the exponent is not
universal and it may depend on the material characteristics as well as on the loading
conditions.

In simple discrete models the scale-free response is associated with a cascade of local
failures. As the external load is slowly increased weak elements fail to hold the imposed
local stress and break. Some internal load redistribution mechanism, whose details depend
on the particular model, increases the local stress on the remaining elements and may
cause further simultaneous local failures. As a result, the material may respond in
avalanches of failure events whose size distribution is often very broad.

For obvious reasons AE is a very useful non-destructive tool that, if properly understood,
may give crucial information about the damage accumulated in loaded materials. Indeed,
mean-field calculations and simulations in some simplified models indicate that the energy
statistics follows a power law with an exponent that changes as the catastrophic failure
point is reached~\cite{hansen05,hansen06}. Events recorded near global failure show a
power law decay with an exponent distinctively smaller than those events recorded far away
from the breakdown point. Interestingly, recent experiments on granite under different
loading conditions have shown a qualitative agreement with this theoretical
observation~\cite{kuksenko}. Should this be a robust and universal property it could be
potentially used as a tool to easily diagnose the damage in loaded materials.

Scaling in fracture is apparently very generic and can also be observed in other
quantities (see for instance Ref.~\cite{alava_zapperi_review} for a recent review). The
distribution of quiet times between AE events also obeys a power law analog of the Omori
law for earthquakes occurrence~\cite{maes,garci2,garci3,alava02,kuksenko,maloy06,rocks07}.
Scale-invariant behavior is also observed in the height-height correlations of fracture
surfaces~\cite{alava_zapperi_review}. The existence of scaling laws in experiments
suggests an interpretation in terms of critical behavior as the global failure point is
approached. However, results are not conclusive and a straightforward link between
experiments and models is still lacking.

From a theoretical point of view fracture physics is a challenging open problem. Materials
are heterogeneous and disorder details may play a crucial role. However, the existence of
power law behavior makes it appealing to try to describe fracture with very simplified
models that include the most relevant symmetries, dimensionality, disorder features, and
loading conditions. 

There has been a great effort devoted to describe fracture crack surfaces in a simple way
by means of stochastic field equations for the dynamics of an advancing front, where the
crack is modeled as an elastic string driven through a disordered medium in the presence
of non-local interactions ~\cite{bouchaud97,fisher98,katzav06,bonamy,katzav07}. However,
continuous crack line models have failed to fully agree with experiments, seemingly
because certain important aspects of fracture, like nucleation of voids in the advancing
front, are not addressed in these continuum approaches. In this respect, lattice models
are expected to be more appropriate. Lattice models describe the elastic solid as a
network of springs with random failure thresholds where the displacement field can be
either scalar or vector. In particular, the random fuse model (RFM)~\cite{arcangelis85}
has attracted a lot of attention as a minimal model for fracture since it is capable of
reproducing some essential features of the fracture phenomenon in a very simple setting.
The RFM represents a scalar electrical analog of the elasticity problem. This
correspondence is based on the formal equivalence between the mathematical form of the
generalized Hooke's law for the scalar elastic problem, $\sigma_{i} = \sum_j g_{ij}
\epsilon_{j}$, and the equations for the electrical problem, $\gamma_{i} =
\sum_j k_ {ij} v_{j}$, where the stress ($\sigma$), local elastic
modulus ($g$) and strain ($\epsilon$) can be mapped to the current density
($\gamma$), local conductance ($k$), and local potential drop ($v$), respectively.
This model goes one step further than mean-field models, like the fiber bundle model, by
including local distribution of stress between the nodes, while preserving the simplicity
in the introduction of disorder as a distribution of thresholds in the system. Much
attention has been devoted to the RFM in the last two decades, including studies
concerning the influence of disorder~\cite{hansen91,hansen91b}, presence of multiscaling
behavior in voltage and current~\cite{arcangelis89}, roughness
exponents~\cite{hansen91,alava00,zap05,alava06,nukala06}, damage
localization~\cite{nukala04}, and failure avalanche
distributions~\cite{hansen94,hansen06,zap97,zap99,zap05,zap05b,zap06}. Interestingly, an
experimental realization of the RFM has recently been studied~\cite{otomar06} to
investigate the influence of disorder on the \textit{I-V} characteristic and crack length.

The dynamics of the RFM takes place in the form of avalanches of failure events.
Therefore, it is tempting to link avalanche activity to the AE observed in experiments.
Several studies have focused on the statistics of avalanches of failure events in the
fuse model~\cite{hansen94,hansen06,zap97,zap99} and improved simulation algorithms have
recently allowed to study larger systems with much better
statistics~\cite{zap05,zap05b,zap06}. However, numerical determinations of the energy
exponent $\beta$ under different loading rates or on different lattices are scarce. 

Lattice models typically predict larger energy exponents $\beta$ than those obtained in
experiments. A number of reasons might be responsible for this discrepancy including the
brittle nature of the models, the existence of dispersion of acoustic waves in
experiments, or the lack of an effective time scale separation between stress relaxation
and loading in real systems, which is assumed in quasi-static models.

Here we study the statistics of the dissipated energy in the RFM under different imposed
strain conditions. We compare seemingly equivalent definitions of the dissipated energy.
For a infinitely slow driving rate (quasi-static model) we find that the PDF of the
released energy shows two different scaling regions. For low energies, the dissipated
energy distribution is robust and independent of the lattice geometry. We find
$\mathcal{P} (E) \sim E^{-1/2}$ and a simple scaling argument is given to explain this
behavior. However, at high energies the distribution exhibits a crossover to a regime
where ${\cal P} (E) \sim E^{-2.75}$, whenever the energy losses are evaluated over the
whole volume of the system. We also study a different estimator of the dissipated energy
that takes into account the energy losses occurring only at microfractures. This
microscopic estimator exhibits strong finite-size effects and fails to show critical
behavior. In a different setting, when the system is driven at finite strain load rates,
we find that the energy distribution decays as $\mathcal{P} (E) \sim E^{-1}$ for all the
energy range.

The paper is organized as follows: In Sec.~\ref{sec:rfm} we define the model, in
Sec.~\ref{sec:ae} we discuss dissipated energy, ways to measure it and we report on our
results for quasi-static and non quasi-static dynamics in the fuse model. Finally, we
conclude with a discussion in Sec.~\ref{sec:disc}

\section{The random fuse model}\label{sec:rfm}
We study a two-dimensional lattice of fuses with unit conductivity $k_{j}=1$ and a
disordered threshold current $\gamma^\mathrm{th}_{j}$ that is a quenched random variable
from a uniform probability distribution in the interval $(0,1)$. An external voltage is
imposed between two bus bars placed at the top and the bottom of the system and periodic
boundary conditions are imposed in the horizontal direction. The system can be driven by a
slowly increasing voltage or current, mimicking the experimental situation where either
strain or stress load can be imposed, respectively. At each update, the Kirchhoff voltage
equations are solved to determine the currents flowing in the lattice. All fuses with
current exceeding the corresponding local threshold are blown and, once the entire system
is below threshold, the voltage (current) is again increased. An avalanche of failure
consists of all the fuses blown between two external voltage (current) increments. Each
fuse behaves linearly until the local current $\gamma_{j}$ reaches the threshold. Then,
the fuse burns and irreversibly becomes insulator $k_{j}=0$, so that burnt fuses are no
longer able to carry any current. The current is redistributed instantaneously after a
fuse is burnt so that current relaxation occurs in time scale that is much faster than the
driving time scale. In our simulations, voltage driving is imposed. The described
configuration should be compared with experimental setups in mode-I fracture where the
strain is slowly increased. An experimental comparison of the AE and critical behavior of
fracture precursors for different load features (strain vs. stress loading) and geometries
has been reported in Ref.~\cite{garci2}. Although imposed strain experiments indicate an
induced plastic deformation in the final stages, no differences were encountered in AE
distribution between the two different loading conditions.

The need to solve a large system of linear equations for each update implies a high
computational cost and limited the reachable system size and the statistical sampling in
this model in the past, when the best performance was achieved by conjugate gradient
methods~\cite{batrouni}. Recently, a new algorithm based on rank-one downdate of sparse
Cholesky factorizations has been introduced~\cite{algorithm1,algorithm2}, which can
largely reduce the computational cost of the simulations. Here we make use of this
algorithm to study two-dimensional networks of fuses in triangular and diamond lattices.
We studied systems of linear size ranging from $L=16$ to $L=256$ and $10^{4}$ realizations
of the disorder.

\section{Dissipated energy}\label{sec:ae}
In stressed materials elastic energy is stored due to redistribution of the external load
all across the lattice. Energy dissipation in the RFM occurs in bursts of breaking events
which can be be compared with the AE observed in experiments. In doing so, one is assuming
that the main contribution to AE is given by the dissipated elastic energy, while
dislocations and friction are, in a slowly driven experiment, secondary contributions to
AE. Also, it is worth to keep in mind that in real systems one expects that only a
fraction of this dissipated energy leads to the AE observed, while the remaining losses
are due to other dissipative mechanisms like dispersion or damping of acoustic waves,
which are not described by purely elastic models like the RFM. Several ways to define the
dissipated energy can be envisaged as we discuss below.

Pradhan {\it et. al.}~\cite{hansen06} found that $\beta = 2.7$ for the RFM on diamond
lattices under stress loading conditions. They calculated the electric power dissipation
in the fuse model as the product of the voltage drop across the network and the total
current that flows through it. The power dissipation in the electrical model is equivalent
to the stored elastic energy in a mechanical system. This is a global definition that
assumes that the whole volume of the system contributes to the dissipated energy.

Alternatively, following Salminen {\it et. al.}~\cite{alava02}, we can define the energy
lost during a given avalanche event $n$ as 
\begin{equation} 
\label{macro_E} 
E_n \sim V_n^2 \Delta G_n \sim V_n^2 s_n, 
\end{equation} 
where $\Delta G_n$ is the change in the elastic modulus due to the failure avalanche,
$s_n$ is the number of broken bonds (avalanche size) of the $n$th event, and $V_n$ is the
corresponding potential drop between the bus bars (strain imposed in the sample).This
definition makes use of the global strain imposed on the system and can be seen as a
coarse-grained or {\em macroscopic} measure of the dissipated energy. 

Note that both, Pradhan {\it et. al.}~\cite{hansen06} and Salminen {\it et.
al.}~\cite{alava02}, definitions of the dissipated energy take into account the whole
volume of the system. However, there is strong experimental evidence indicating that AE is
actually a localized phenomenon in space and time so that energy release actually occurs
at microfractures~\cite{garci1,garci2}, and it is not therefore spread across the system.
This suggests one should consider other ways to define the dissipated energy in the model,
in particular, it may be interesting to study measures of released energy that are
directly linked to the bonds involved in avalanches of local breaking events.
\begin{figure}
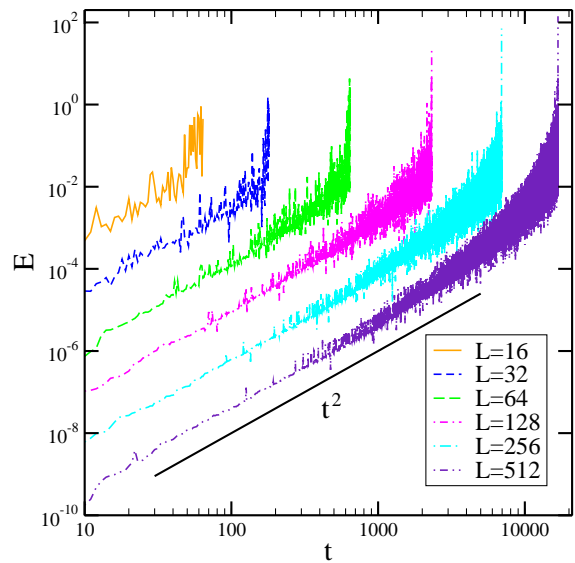

\centerline{\includegraphics *[width=75mm,type=eps,ext=.eps,read=.eps]{fig1}}
\caption{(Color online) Temporal evolution of microscopic energy until breakdown for a
typical realization of the disorder on diamond lattices ranging from $L=16$ (top curve) to
$L=512$ (bottom curve). A first region with slope $\alpha=2$ is followed by a second
region dominated by fluctuations.}
\label{fig1}
\end{figure}

In this spirit, we introduce here what we call {\em microscopic} dissipated energy, which
is defined as the sum of the energy losses at every element of the system involved in the
$n$th failure avalanche. This can be calculated by adding up the energy dissipated by each
individual broken bond, $\gamma_{j}=k_{j}v_{j}^{2}$, where $v_{j}$ is the local potential
drop at bond $j$. Since fuses break right at the threshold we can define the {\em
microscopic} dissipated energy due to the $n$th failure avalanche as
\begin{equation}
\label{micro_E}
E_{n}=\sum_{j}^{{s}_{n}} (\gamma^\mathrm{th}_{j})^2,
\end{equation}
where the sum runs over each broken bond within the $n$th avalanche. 

We devote the rest of the paper to analyze the dissipated energy statistics in the fuse
model for the case of infinitesimal strain (quasi-static model) and also under different
finite strain rates (non quasi-static model). We shall be comparing the numerical results
for the different measures of the dissipated energy discussed above for triangular and
diamond lattices. One would expect that the three definitions give similar temporal
behavior for the dissipated energy statistics, apart from constant factors. Although this
is actually the case in the low energy range, dissipation statistics differs at high
energies for different estimators.

\subsection{Quasi-static case: infinitesimal strain rate}
Let us first focus on infinitesimal driving. For the sake of computational convenience, a
voltage $V=1$ is fixed between the two bus bars. The next fuse to burn is determined by
$\mbox{max}_{j}(\gamma_{j}/\gamma^\mathrm{th}_{j})$ so the external voltage is increased
until this fuse exactly reaches its threshold. The new current configuration is calculated
according to Kirchhoff equations. This rearrangement can cause other fuses to overpass
their thresholds without further voltage increase. All the fuses burnt at the same
external voltage constitute an avalanche. This process is repeated until the network
becomes disconnected.
\begin{figure}
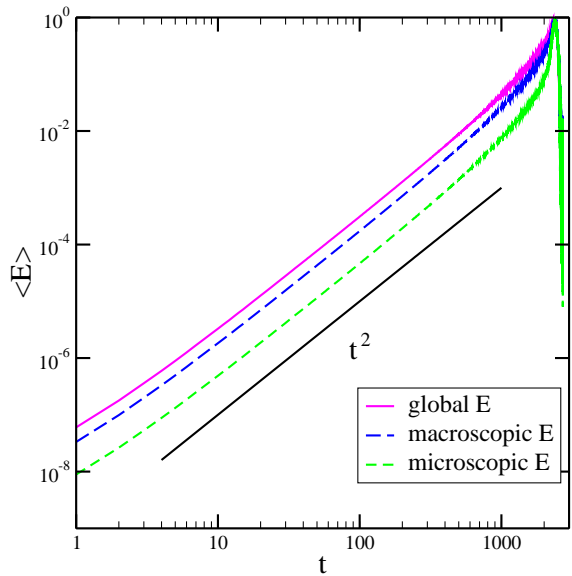

\centerline{\includegraphics *[width=75mm,type=eps,ext=.eps,read=.eps]{fig2}}
\caption{(Color online) Temporal evolution of energy dissipation averaged over $10^{4}$
realizations in a $L=128$ diamond lattice. All the three energies, global, macroscopic and
microscopic, exhibit the same $t^{2}$ trend. The three curves have been rescaled by its
maximum for clarity.}
\label{fig2}
\end{figure}

Quasi-static dynamics results in two very well separated time scales, {\i.e.} a fast
relaxation process and a slow external driving. As occurs in other systems
exhibiting a well defined time scale separation (as for instance in self-organized
criticality), the natural time scale is given by the slow time scale. Therefore, in the
following, time refers to the number of avalanches occurred. In Fig.~\ref{fig1} we show a
typical realization of the temporal evolution for the dissipated energy according to
Eq.~(\ref{micro_E}) for different system sizes.  One can see that for each realization the
released energy grows in time as a power-law $E(t) \sim t^{2}$. At later times,
fluctuations around the trend increase (the larger the bigger the system size) as the
system approaches the total breakdown point. This dynamic behavior is very robust and
independent of the system size or lattice type. 

We also found identical temporal behavior for the macroscopic dissipated energy,
Eq.~(\ref{macro_E}), and for the dissipated electric energy used by Pradham {\it et. al.}
for both diamond and triangular lattices. In Fig.~\ref{fig2} we compare the temporal
evolution of the average dissipated energy on the diamond lattice according to the three
definitions. Notably, this behavior is also in agreement with that reported in
Ref.~\cite{zapDynamic} for a very different dynamical spring model that included acoustic
waves. The origin of the robust $t^2$ growth law is perhaps more transparent in
Eq.~(\ref{macro_E}), where the driving potential is increasing linearly with time (as we
are imposing a quasi-static dynamics).

The temporal power-law trend gives significant information about the functional form of
the dissipated energy statistics in the RFM. The probability density function (PDF) of the
dissipated energy at time $\tau$ in a system of lateral size $L$ is given by ${\cal
P}_{\tau}(E,L)=(1/\tau)\int_{0}^{\tau} {dt\,\delta[E(t,L)-E]}$, where $\delta(u)$ is the
Dirac delta distribution. This corresponds to the energy distribution to be observed after
the first $\tau$ avalanche events. Note that the probability distribution ${\cal
P}_{\tau}(E,L)$ is expected to be non-stationary, so it should depend explicitly on the
observation time $\tau$. Let us now consider that the dissipated energy grows in time as a
power-law with some exponent, $E(t,L)\sim{L^{-\sigma}\,t^{\alpha}}+\eta(t)$, where
$\sigma$ captures the scaling with system size observed in Fig.~\ref{fig1} and $\eta(t)$
is a noise term representing the random fluctuations around the trend. The details of the
noise term $\eta(t)$ are not known, but one can argue they may depend non-trivially on the
interplay between the evolving currents and the disordered thresholds. Actually, as can be
readily seen in Fig.~\ref{fig1}, fluctuations are strongly asymmetric around the average,
which immediately implies a non-Gaussian distribution of $\eta$, possibly including
non-trivial correlations. Despite these difficulties one can perform the integral in
certain limit, up to certain energy cut-off $E_\times$ below which fluctuations of the
energy are negligible. We have 
\begin{eqnarray}
\label{P_E}
\lefteqn{\mathcal{P}_{\tau}(E,L)= \tau^{-1} \int_{0}^{\tau}
{dt\,\delta[L^{-\sigma}\,t^{\alpha} + \eta(t) - E]} =} \nonumber\\
& & = \tau^{-1} L^{\sigma/\alpha}\,E^{-1+1/\alpha} \times \nonumber\\
& & \times \int_{0}^{\tau/{(L^{\sigma}E)}^{1/\alpha}}ds \, \delta[s^{\alpha} +
\eta(L^{\sigma/\alpha}E^{1/\alpha}s)/E - 1],
\end{eqnarray}
and thus keeping only the lowest-order term we arrive at
\begin{equation}
\label{PTE}
\mathcal{P}_{\tau}(E,L)\sim{ \tau^{-1}L^{\sigma/\alpha}E^{-1+1/\alpha}}
\end{equation}
for $\tau/{(L^{\sigma}E)}^{1/\alpha} \gg 1$. This immediately leads to the existence of a
characteristic energy scale $E_\times \sim L^{-\sigma}\tau^\alpha$ above which energy
fluctuations dominate the statistics. It is clear that the details of the noise statistics
(including the distribution and temporal correlations) would be required to obtain the
specific mathematical form of the dissipated energy distribution above the characteristic
energy $E_\times$. For the RFM we have an algebraic growth with exponent $\alpha \approx
2$ (see Fig.~\ref{fig1}), so we expect to have a energy distribution decaying as $\sim
E^{-1/2}$ for energies $E < E_\times$.
\begin{figure}
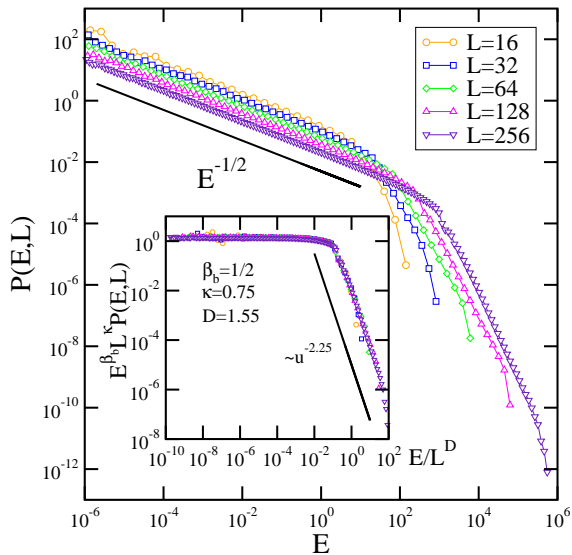

\centerline{\includegraphics *[width=75mm,type=eps,ext=.eps,read=.eps]{fig3}}
\caption{(Color online) Probability distribution of the global dissipated energy for
different system sizes. The low energy region decays as a power-law with exponent $\beta =
1/2$ and shows a crossover at $E_\times$. The inset shows data collapse according to
Eq.~(\ref{FSS}). The values of $\kappa$ and $D$ are in good agreement with the expected
relation $D=\alpha\kappa$. Logarithmic binning has been employed.}
\label{fig3}
\end{figure}

We are interested here in the distribution statistics after complete breakdown is
attained. The characteristic time to total failure is expected to scale with system size
as $T_\mathrm{break} \sim L^{z}$, where $z$ is the dynamic exponent. From Eq.~(\ref{PTE})
the energy statistics after failure, $\mathcal{P}(E,L) \equiv
\mathcal{P}_{\tau=T_\mathrm{break}}(E,L)$, reads
\begin{equation}
\label{PTE_s}
\mathcal{P}(E,L)\sim{ L^{-z+\sigma/\alpha}E^{-1+1/\alpha}},
\end{equation}
for energies below a crossover energy $E_\times \sim L^{\alpha\,z-\sigma}$.

\begin{figure}
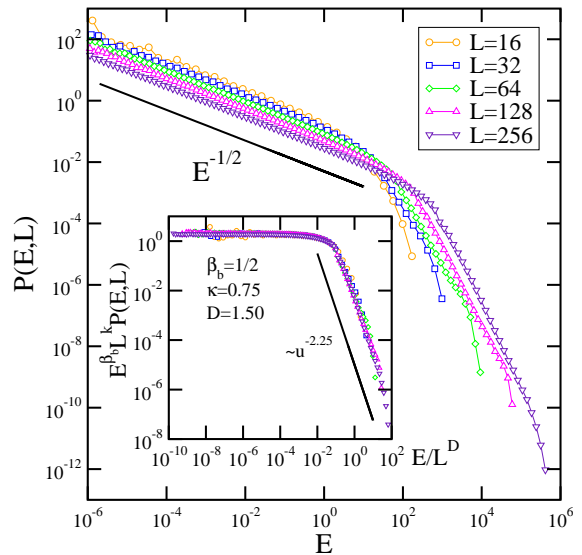

\centerline{\includegraphics *[width=75mm,type=eps,ext=.eps,read=.eps]{fig4}}
\caption{(Color online) Probability distribution of macroscopic dissipated energy,
Eq.~(\ref{macro_E}), for different system sizes and the corresponding data collapse
(inset). The same behavior and exponents as in the global definition are found} 
\label{fig4}
\end{figure}

Figure~\ref{fig3} shows the probability distribution, $\mathcal{P}(E,L)$, with statistics
collected up to total failure for the above introduced global dissipated energy. Two
regions can be readily distinguished. The low energy statistics is in excellent agreement
with a power-law decay $\sim E^{-1/2}$ over several decades in energy. Similar behavior is
observed in Fig.~\ref{fig4} for the macroscopic energy measure defined in
Eq.~(\ref{macro_E}). 

The dependence with system size of the numerical data observed in Figs.~\ref{fig3} and
\ref{fig4} can be better characterized by means of a finite-size scaling analysis. The
behavior of the energy distribution suggests the scaling ansatz
\begin{equation}
\label{FSS} \mathcal{P}(E,L)\sim{E^{-\beta_\mathrm{b}} L^{-\kappa}
\mathcal{G}(E/E_\times)},
\end{equation}
where the scaling function $\mathcal{G}(u)\sim \mathrm{const}$ for $u \ll 1$
and becomes $\mathcal{G}(u) \sim u^{\beta_\mathrm{b} - \beta_\mathrm{a}}$ for $u \gg 1$.
$\beta_\mathrm{a}$ and $\beta_\mathrm{b}$ are the scaling exponents of the distribution
above and below the crossover, respectively. The crossover energy scales with system size
as $E_\times \sim L^{D}$ with some critical exponent $D$.

We can now make use of the theoretical relation we derived in Eq.~(\ref{PTE_s}) to prove
that the two scaling exponents $\kappa$ and $D$ are not independent. Comparing
Eqs.~(\ref{FSS}) and (\ref{PTE_s}) one obtains that the following scaling relations must
be fulfilled:
\begin{eqnarray}
D &=& \alpha z - \sigma \nonumber\\
\kappa &=& z - \sigma/\alpha, \nonumber\\
\beta_\mathrm{b} &=& 1 - 1/\alpha
\label{scal_rel}
\end{eqnarray}
which immediately imply that $D = \alpha \kappa$. Also, according to our estimate 
$\alpha = 2$ from Fig.~\ref{fig1} we should have $\beta_\mathrm{b} =1/2$. This reduces
the number of free exponents to achieve a good data collapse~\footnote{We can also
determine the dynamic exponent $z=1.75(3)$ by counting the average number
of avalanches taking place before total failure in a system of lateral size $L$ which
should scale as $T_\mathrm{break} \sim L^{z}$. It is interesting to note that the
specific value of $z$ is not required to produce the data collapse in Eq.~(\ref{FSS}).}.

The insets of Figs.~(\ref{fig3}) and (\ref{fig4}) show a data collapse according to
Eq.~(\ref{FSS}) with exponents $\kappa = 0.75(4)$, $D= 1.55(5)$ and $\kappa = 0.75(4)$,
$D= 1.50(5)$, respectively, and the energy exponent below the crossover $\beta_\mathrm{b}
= 1/2$. The fit of the scaling function for $u \gg 1$ corresponds to the difference
$\beta_\mathrm{b} - \beta_\mathrm{a} = - 2.25(2)$, and implies that the scaling exponent
of the energy distribution above the crossover is $\beta_\mathrm{a} = 2.75(2)$, identical
within error bars for both energy measures. This exponent is to be compared with the one
calculated by Pradham {\it et. al.} for the electric power dissipation in the high energy
region for the diamond lattice in Ref.~\cite{hansen06}, where they report $\beta=2.7$ over
two decades of energy. The macroscopic energy defined in Eq.~(\ref{macro_E}) not only
exhibits the same behavior but the same exponents in the two regions indicating both
definitions are completely equivalent. However, as we show below this is not the case for
the microscopic energy statistics.

\begin{figure}
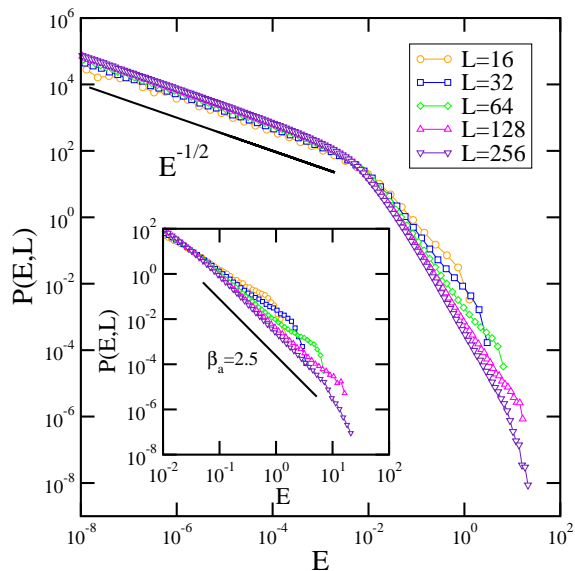

\centerline{\includegraphics *[width=75mm,type=eps,ext=.eps,read=.eps]{fig5}}
\caption{(Color online) Probability distribution of microscopic dissipated energy,
Eq.~(\ref{micro_E}), for different system sizes. The inset shows a zoom of the high-energy
region which decays with an exponent around $\beta = 2.5$ over two decades for the largest
system. Data show strong finite-size effects and do not obey good scaling behavior.} 
\label{fig5}
\end{figure}

Figure~\ref{fig5} shows the behavior of the microscopic energy defined in
Eq.~(\ref{micro_E}). Recall that this measure is intended to collect only those
contributions to the released energy coming from sites participating in the failure
avalanche. We find that in the low-energy region the distribution also decays as
$\mathcal{P}(E,L) \sim E^{-1/2}$. However, in this case we observe that the probability
does not seem to depend significatively on system size, $\kappa \approx 0$.
Correspondingly, $D = \alpha \kappa \approx 0$ and the crossover energy $E_\times$ does
not vary with system size. The lack of system size dependence of the microscopic energy
may be related to the fact that the macroscopic and global estimators are sensible to the
whole volume of the system, while the microscopic energy is not. The inset of
Fig.~\ref{fig5} shows a zoom of the high energy region, where strong finite-size effects
are demonstrated by the variation of the exponent $\beta_\mathrm{a}$ with system size. Data
in Fig.~\ref{fig5} obviously fail to exhibit finite-size scaling.

It is worth to stress here that the PDF for all the three energy definitions exhibits
identical scaling behavior at low energies, $\sim E^{-1/2}$, and is robust to changes
in system size and lattice type. This universality arises from the $t^2$ growth law of the
dissipated energy which is a feature shared by all the three definitions for any system
size and lattice geometry. However, the lack of scaling behavior with system size of the
microscopic energy has a direct impact on the high energy regime of its probability
distribution.

\subsection{Non-stationarity and signatures of imminent failure} The temporal series of
the dissipated energy in the fuse model are highly non stationary, as can be easily
noticed in Fig.~\ref{fig1}, and so the probability density $\mathcal{P}_\tau(E,L)$ depends
explicitly on the observation time $\tau$. This has been claimed to be useful to signal
the onset of catastrophic failure~\cite{hansen05,hansen06}, with evident practical
applications for diagnosing damage in loaded materials. 

In ~\cite{hansen06} the power dissipation avalanche distribution for the entire breakdown
process was compared with that obtained only in a very narrow window around breakdown. In
order to do this, those authors first computed the average over disorder samples of the
number of fuses $\langle N_\mathrm{break}\rangle$ blown before catastrophic failure and,
for every realization, collected statistics from events after almost $\langle
N_\mathrm{break} \rangle$ fuses have blown. A possible drawback of that procedure is that,
since the time required to reach total failure largely varies among different disorder
samples, one is mixing realizations that are very close to complete failure with others
that are, say, half way into it, which obey a different statistics.

Our procedure to obtain the statistics differs significantly from that used in
Refs.~\cite{hansen05,hansen06} and has the advantage that it is not affected by this
undesired effect. Moreover, in contrast to ~\cite{hansen05,hansen06} we want to compare
here the distribution of released energy until breakdown with that obtained when the
system is at the very beginning of its evolution and how it changes as we approach
failure.  We proceed as follows. For each disorder realization we let the system evolve up
to total breakdown, which gives the corresponding $T_\mathrm{break}$ for that
particular disorder realization. We then compare the collected statistics with that
observed for that particular disorder realization up to two intermediate times, $\tau =
T_\mathrm{break}/8$ and $T_\mathrm{break}/2$, that is, with the probability density when
only the first one eighth and half of the failure avalanches are counted, respectively.
So that we collect statistics from realizations at the same evolution stage.
\begin{figure}
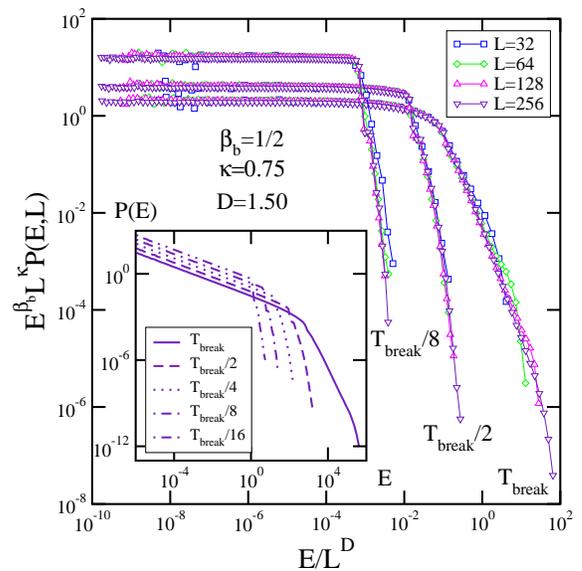

\centerline{\includegraphics *[width=75mm,type=eps,ext=.eps,read=.eps]{fig6}}
\caption{(Color online) Macroscopic energy distribution for different
final observation times. The same exponents collapse the distribution independently 
of the observation time but the slope of the high-energy tail is shifted to larger values 
when one measures further away from the breakdown instant. The inset shows the energy 
distribution for more intermediate observation times for $L=256$. It can be clearly
observed the slope change at breakdown.}
\label{fig6}
\end{figure}

For each observation time, the probability density $\mathcal{P}_\tau(E,L)$ decays as $\sim
E^{-1/2}$ until the crossover energy $E_\times(\tau,L)$. In Fig.~\ref{fig6} we plot the
behavior of the distribution for the macroscopic energy, Eq.~(\ref{macro_E}), for
different system sizes and different observation times, rescaled according to
Eq.~(\ref{FSS}). It can be observed that, while the crossover shifts to larger energy
values as we approach complete breakdown, the scaling exponent in the second region is
conserved as we increase the observation times. However, an abrupt change of exponent is
observed only at the breakdown time $T_\mathrm{break}$. This effect is perhaps better
visualized in the inset of Fig.~\ref{fig6} where we plot the distribution data (unscaled)
for the largest system $L=256$ and different observation times. Despite we used a
different measure of the dissipated energy and a different way to collect events these
results are in agreement with the crossover picture between the two limiting behaviors
reported in Refs.~\cite{hansen05,hansen06}. This indicates that non-stationary effects of
the dissipated energy temporal signal may actually be useful to characterize damage in
stressed materials.

\subsection{Finite driving rate} The lack of time scale separation in real experiments has
been suggested \cite{alava_zapperi_review} as a possible reason for the discrepancy with
the typical exponents found in numerical simulations of quasi-static models. In order to
investigate this point we have studied the RFM under finite driving rates, so that the
model evolution is no longer quasi-static. We have analyzed both stress and strain loading
conditions and our results were not affected by the loading mode we used. For the sake of
brevity here we only report on our results for the latter.

Strain is applied on the system by imposing a small potential drop between the bus bars in
such a way that all the fuses are initially below threshold. The voltage is then increased
at a fixed rate $dV/V$ letting all the fuses over threshold burn, instead of the slow
driving setup studied above. As before, an avalanche is defined as all the fuses burnt
between two consecutive voltage increments. In this setup, we can still observe an
effective time scale separation at the early stages of the evolution, while potential
increments are small.
\begin{figure}
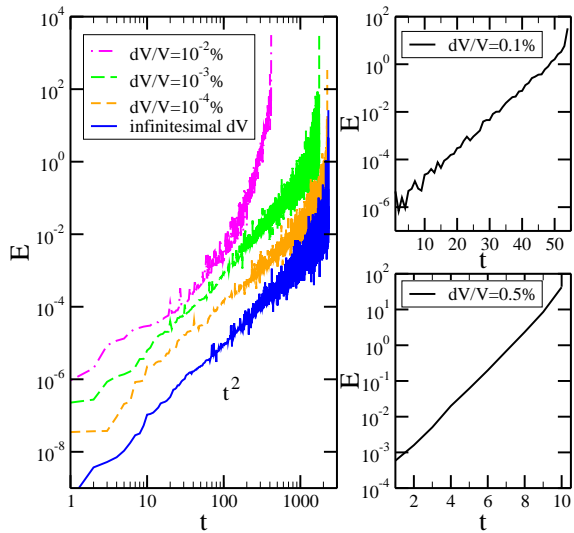

\centerline{\includegraphics *[width=75mm,type=eps,ext=.eps,read=.eps]{fig7}}
\caption{(Color online) Temporal evolution of microscopic energy in an $L=128$ diamond
lattice for several finite driving rates. On the left panel, it can be observed how, as
the driving is increased, the energy starts to deviate from $t^2$ growth. On the right
panels, one can see that, for large enough driving, the energy grows exponentially in
time. Note the linear-log scale in the latter.}
\label{fig7}
\end{figure}

Figure~\ref{fig7} summarizes our numerical results for the time evolution of the
microscopic dissipated energy, Eq.~(\ref{micro_E}), at different strain rates in diamond
lattices of linear size $L = 128$. We observe that a power-law trend $E  \sim t^{\alpha}$
is satisfied with a larger exponent $\alpha$ as the strain rate increases. If the strain
rate is large enough (about $dV/V= 0.01 \%$) the evolution is no longer described by a
power-law, but it becomes exponential in time. The same behavior is observed if either
the macroscopic or global energy are used.

According to our simple calculation in Eq.~(\ref{PTE}), we expect that, in the limit of
exponential growth, $\alpha \gg 1$, the PDF of the dissipated energy becomes
$\mathcal{P}_\tau (E) \sim E^{-1}$ for $E \ll E_\times$. We observe that the crossover
energy $E_\times$ diverges as the strain rate is increased. In Fig.~\ref{fig7} one can
clearly see that when we increase the strain rate the energy fluctuations become much
smaller. In fact, fluctuations are negligible for the whole temporal (energy) range for
large enough driving rates, when the exponential growth sets in (see Fig.~\ref{fig7}).
This means that the fall-off tail of the distribution corresponding to energies $E \gg
E_\times$ is completely washed out in the case of large enough strain rates. Therefore we
expect the dissipated energy probability to be
\begin{equation}
\label{PTE_finite}
\mathcal{P}(E) \sim E^{-1}
\end{equation}
if we let the system evolve up to complete breakdown, $\tau = T_\mathrm{break}$, as well
as for any other intermediate times.
\begin{figure}
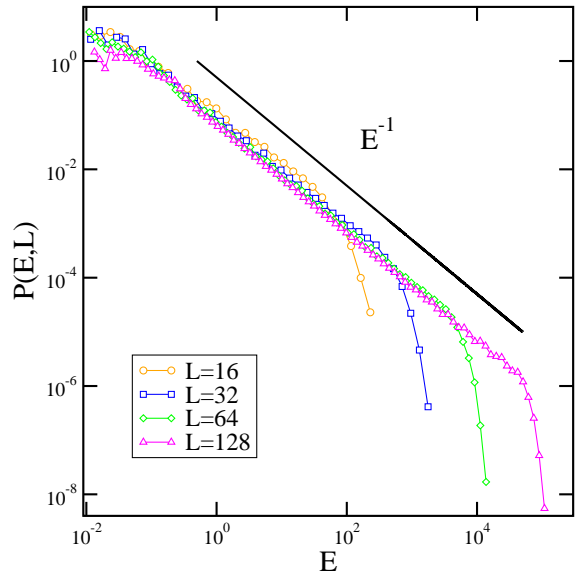

\centerline{\includegraphics *[width=75mm,type=eps,ext=.eps,read=.eps]{fig8}}
\caption{(Color online) Macroscopic energy distribution for $L=16,32,64$ and $128$ diamond
lattices
under finite strain increments $dV/V=10^{-1}\%$. A single power-law regime with the
predicted exponent $\beta = 1$ is observed over the whole range of energies up to a
finite-size cutoff.}
\label{fig8}
\end{figure}

In Fig.~\ref{fig8} we show the PDF of the macroscopic dissipated energy under a strain
loading rate $dV/V = 0.1\%$ measured for all events up to total failure in the diamond
lattice for different system sizes. Identical results are obtained for the global
energy estimator (not shown). Our numerical results are in excellent agreement with
the prediction for finite loading rates in Eq.~(\ref{PTE_finite}). The probability
distribution for the microscopic energy also scales as $E^{-1}$ for the whole range of
energies in the case of finite-driving (not shown), but the crossover energy shows no
dependence with system size, in agreement with our above discussed results $D \approx 0$.

The conclusion is that the lack of time scale separation leads to a significative change
in the distribution of the dissipated energy in the fuse model, as it should be expected.
From a physical point of view there are strong differences in the system dynamics in the
case of infinitesimal driving as compared with finite driving. If the system is driven at
finite rates, relaxation to one of the infinitely many metastable configurations is not
reached before a new perturbation acts on the system. This gives rise to a highly
nonlinear superposition of cascades of released energy instead of individual well-defined
avalanche events. A finite driving rate generically leads to a growth of the dissipated
energy at a much faster rate than the usual quasi-static dynamics, possibly exponential
for any finite driving rate in large enough systems. In turn, this fast growth takes the
crossover energy $E_\times$ to exponentially large values. The result is that the range of
energies in which the PDF of the dissipated energy is described by Eq.~(\ref{PTE_finite})
becomes very large, actually covering the whole range of energies. 

\section{Discussion}\label{sec:disc}
We have studied the RFM for fracture under strain loading conditions. We have focused on
the dissipated energy in order to test the 
validity of the model to account for the AE statistics observed in real 
experiments in loaded materials. Different ways to define the released energy have been
discussed, including a microscopic quantity that takes into account just the energy losses
at each broken bond during an avalanche. 

Our results indicate that, for quasi-static dynamics, the dissipated energy statistics
exhibits two very different regions depending on the energy scale one is looking at. These
two scaling regions are separated by a typical energy $E_\times \sim
L^{\alpha\,z-\sigma}$ and obey finite-size critical behavior. The low-energy
region, for $E < E_\times$, is well described by a power-law decay $\mathcal{P}(E)\sim
E^{-1/2}$, which is robust and independent of lattice geometry. We gave a simple scaling
argument showing that this robustness is linked to the generic $t^2$ growth law of the
dissipated energy; a feature shared by all the energy estimators we studied on any system
size and lattice type and that it is directly linked to the quasi-static nature of the
model. The statistics above the typical energy $E > E_\times$ crosses over to
$\mathcal{P}(E)\sim E^{-2.75}$ and ranges several decades in energy. 

Apart from scaling factors, the three energy definitions used here were expected to show
similar statistics. However, while the behavior of macroscopic and global energies can be
captured by the same scaling exponents and the high-energy region exponent is well
defined, this is not the case for the microscopic energy that, although it obeys the same
scaling form, shows no system size dependence and a different (size-dependent) exponent
for the high energy region is obtained. Regarding the microscopic energy we introduced
here one must admit that scaling in the high-energy region is not satisfying in general
terms. Not only the numerical value of the exponent depends on the lattice size and fails
to exhibit finite-size scaling, but also the scaling region covers a very narrow energy
range. This should be particularly relevant when comparing with real fracture experiments
that have shown that dissipated energy participating in AE is not released all across the
sample, but, quite the opposite, localized at microfractures~\cite{garci1,garci2}. 

Finally, we also studied the fuse model at finite driving rates. It is an often expressed
belief that relaxing the quasi-static condition might lead to $\beta$ exponents that
compare better with experiments. We showed that under finite driving the cut-off energy
diverges exponentially, so that the scaling $\mathcal{P} (E) \sim E^{-1}$ dominates all
the energy range at any given time for large enough driving rates (possibly for any finite
driving rate in large enough systems). The conclusion is that relaxing the quasi-static
condition cannot give account of experiments, where $\beta$ typically ranges between $1.2$
-- $2.0$ depending on the material.

The evidence we have up to now about the RFM indicates that it might well be the case that
other essential aspects to quantitatively account for AE energy exponents in real
materials, like plasticity, dislocations, friction, damping of acoustic waves, etc., are
missing in the admittedly oversimplified fuse model.

\acknowledgments 
Financial support from the Ministerio de Educaci\'on y Ciencia (Spain) under project
FIS2006-12253-C06-04 is acknowledged. CBP is supported by a FPU fellowship (Ministerio de
Educaci{\'o}n y Ciencia, Spain).

\end{document}